\documentclass[a4paper,11pt]{article}
\usepackage{pos}

\newcommand{\me}{m_{\rm e}}
\newcommand{\mi}{m_{\rm i}}

\newcommand{\wi}{\omega_{\rm pi}}

\newcommand{\Rw}{\mathcal{R}_{\rm w}}
\newcommand{\Rs}{\mathcal{R}_{\rm sh}}

\title{Electron-Ion Temperature Ratio in\\ Transrelativistic Unmagnetized Shock Waves}

\author*[a,b,c]{Arno Vanthieghem}
\author[b]{Vasileios Tsiolis}
\author[d]{Frederico Fiuza}
\author[c]{Kazuhiro Sekiguchi}
\author[b]{Anatoly Spitkovsky}
\author[e]{Yasushi Todo}

\affiliation[a]{Sorbonne Université, Observatoire de Paris, Université PSL, CNRS, LERMA,\\
Paris F-75005, France}

\affiliation[b]{Department of Astrophysical Sciences, Princeton University,\\
Princeton 08544, New Jersey, USA}

\affiliation[c]{Department of Astro-fusion Plasma Physics (AFP), Headquarters for Co-Creation Strategy,\\
Tokyo 105-0001, Japan}

\affiliation[d]{GoLP/Instituto de Plasmas e Fusão Nuclear, Instituto Superior Técnico, Universidade de Lisboa\\
Lisbon 1049-001, Portugal}

\affiliation[e]{National Institute for Fusion Science, National Institutes of Natural Sciences, Toki,\\
Gifu 509-5292, Japan}

\emailAdd{arno.vanthieghem@obspm.fr}

\abstract{Weakly magnetized shock waves are paramount to a large diversity of environments, including supernova remnants, blazars, and binary-neutron-star mergers. Understanding the distribution of energy between electrons and ions within these astrophysical shock waves spanning a wide spectrum of velocities is a long-standing challenge. In this study, we present a unified model for the downstream electron temperature within unmagnetized shock waves. Encompassing velocities from Newtonian to relativistic, we probe regimes representative of the gradual deceleration of the forward shock in the late gamma-ray burst afterglow phase, such as GRB 170817A. In our model, heating results from an ambipolar electric field generated by the difference in inertia between electrons and ions, coupled with rapid electron scattering in the decelerating turbulence. Our findings demonstrate that the electron temperature consistently represents $10\%$ of the incoming ion kinetic energy in the shock front frame over the full range of shock velocities.}

\FullConference{%
  High Energy Phenomena in Relativistic Outflows VIII (HEPROVIII)\\
  23-26 October, 2023\\
  Institut d'Astrophysique de Paris, Paris, France
}

\begin{document}
\maketitle

\section{Introduction}

Weakly magnetized collisionless shock waves efficiently heat electrons well above $T_{\rm e}/T_{\rm i}\,\sim\,\me/\mi$ inferred from pure adiabatic compression. In the non-relativistic regime, $|v_{\rm sh}|\,\ll\,c$, with $v_{\rm sh}$ the shock velocity in the upstream frame\footnote{Except if explicitly mentioned, quantities are defined in the shock-front frame $\mathcal{R}_{\rm s}$. The shock velocity $v_{\rm sh}\,=\,v_{\rm sh|u}$, measured in the upstream frame, relates to the four-velocity normalized to the speed of light $u_{\rm sh}\,=\,\gamma_{\rm sh} \beta_{\rm sh}$ with $\beta_{\rm sh}\,=\,v_{\rm sh}/c$ and $\gamma_{\rm sh}\,=\,1/\sqrt{1 - \beta_{\rm sh}^2}$.}, a temperature ratio of the order of $T_{\rm e}/T_{\rm i}\,\sim\,0.3$ is extracted from simulations. Likewise, the relativistic regime, $|v_{\rm sh}|\,\lesssim\,c$, suggests $T_{\rm e}/T_{\rm i}\,\sim\,0.5$. Observations support strong nonadiabatic heating, $T_{\rm e}/T_{\rm i}\,\sim\,\mathcal{O}(0.1)$, over the full range of shock velocities, as inferred from in situ measurement at Earth's bow shock~\cite{Feldman_1982, Johlander_2023}, radio and X-ray synchrotron emissions from young Supernova Remnants~\cite{Reynolds_2008} (see~\cite{Ghavamian_2013, Raymond_2023} and references therein for further details), to the modeling of ultrarelativistic gamma-ray burst afterglow emission~\cite{Freedman_2001} with important implications on the observational signature~\cite{Vanthieghem_2020}. 

Understanding and modeling the heating of electrons is essential to describing their injection process into diffusive shock acceleration and interpreting and modeling their radiative spectra. Here, we focus on providing a unifying picture of the electron-ion temperature ratio in transrelativistic shock waves mediated by a microturbulent magnetic field.

\section{Kinetic simulations}

\begin{figure*}[ht]
	\begin{center}
	\includegraphics[width=0.85\textwidth]{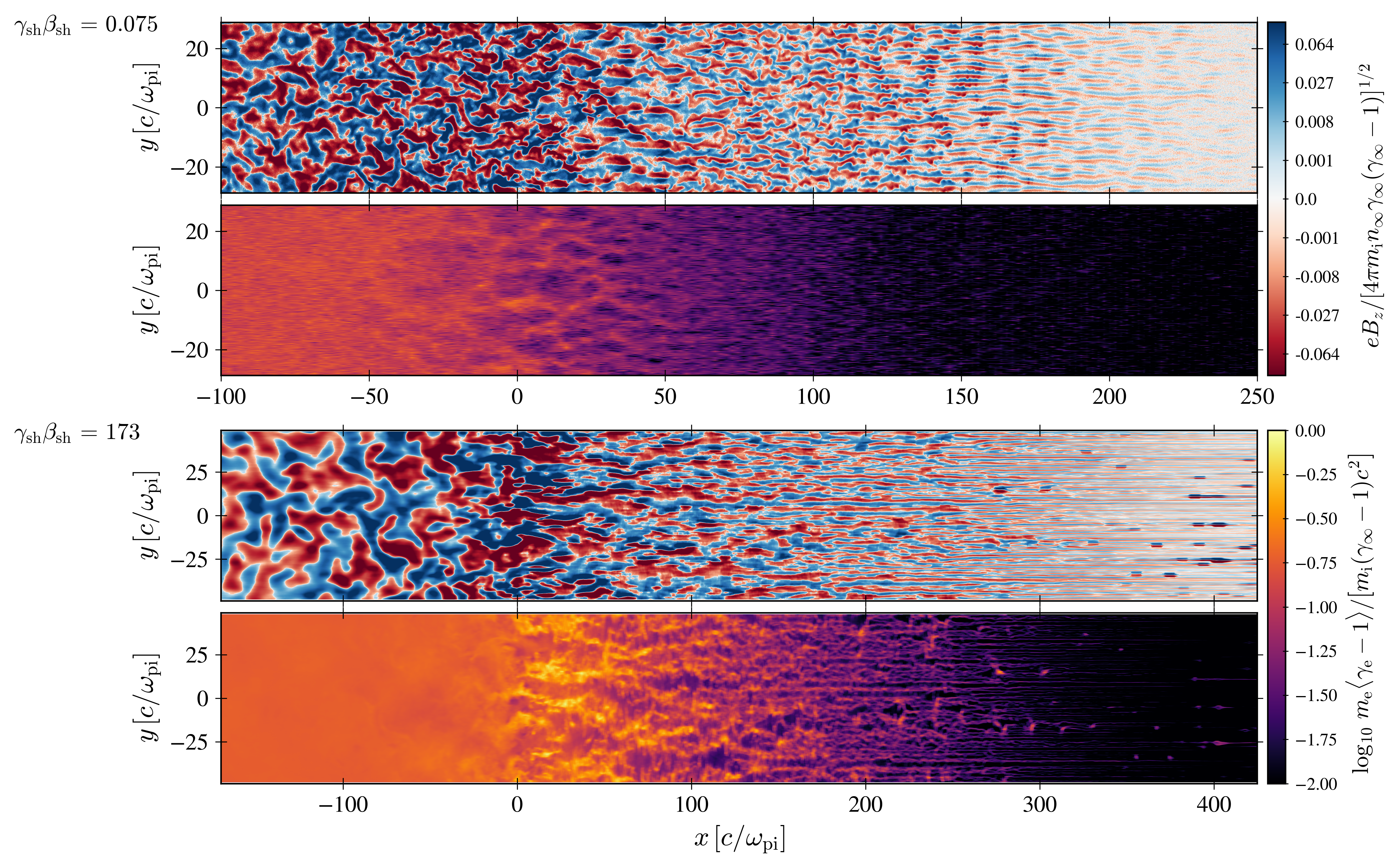}
	\caption{Closeup of the precursor of unmagnetized electron-ion collisionless shocks with $u_{\rm sh}\,=\,0.075$ (top) and $u_{\rm sh}\,=\,173$ (bottom). The respective top and bottom panels in each case correspond to the magnetic field amplitude normalized by the square root of the incoming momentum flux and the average kinetic energy of the electrons normalized by the incoming kinetic energy of the ions. The approximate shock transition is located at $x\,=\,0$. The figure is adapted from~\cite{Vanthieghem_2022,Vanthieghem_2024}.}
	\label{fig:prof_ga100}
	\end{center}
\end{figure*}

Electron energization in weakly magnetized shocks is a nonlinear process requiring the self-consistent modeling of the electron-ion interaction through a self-generated microturbulent electromagnetic field. To illustrate the process, we first discuss a series of Particle-In-Cell (PIC) simulations covering an extensive range of velocities from non-relativistic to ultra-relativistic. Figure~\ref{fig:prof_ga100} illustrates the magnetic field profile and bulk energy of the electrons in the non-relativistic and relativistic cases. These kinetic simulations are performed in a 2D3V geometry (2D in physical space, 3D in momentum space) using the finite-difference time-domain, PIC code \textsc{Tristan}~\citep{Anatoly_2005} and \textsc{calder}~\citep{Lefebvre_2003}. To ensure the stability of the plasma over the shock crossing time, we use two different configurations for the non-relativistic and relativistic shock regimes. Newtonian shock simulations are performed with \textsc{Tristan} for four-velocities between $|u_{\rm sh|u}|\,=\,|\beta_{\rm sh|u}\gamma_{\rm sh|u}|\,\simeq\,[0.075, 0.2]$ and mass ratio $\mi/\me\,=\,49$. Numerical heating is significantly reduced by imposing the upstream plasma to be at rest, and a reflecting conducting piston moves from the left with positive velocity. Such a setup is not applicable to the relativistic regime due to the strong compression of the downstream in the upstream frame. Hence, the simulation is run in the downstream frame, and we filter the relativistic beam-grid instability using the Cole-Karkkainen electromagnetic solver~\citep{Karkkainen_2006} coupled with the Godfrey-Vay filtering method~\citep{Godfrey_2014}. This approach allows for large current factor $c \Delta t/\Delta x\,=\,0.99$ in 2D. In this regime, we discuss shock simulations with $|u_{\rm sh|u}|\,=\,|\gamma_{\rm sh|u} \beta_{\rm sh|u}|\,\simeq\,[17, 175]$ and 
mass ratio $\mi/\me\,=\,[25, 100]$. In both cases, the particles are injected from the left. Table~\ref{tab:simulations} enumerates the simulations and parameter ranges probed.

\begin{figure}[!ht]
    \centering
    \includegraphics[width=0.65\textwidth]{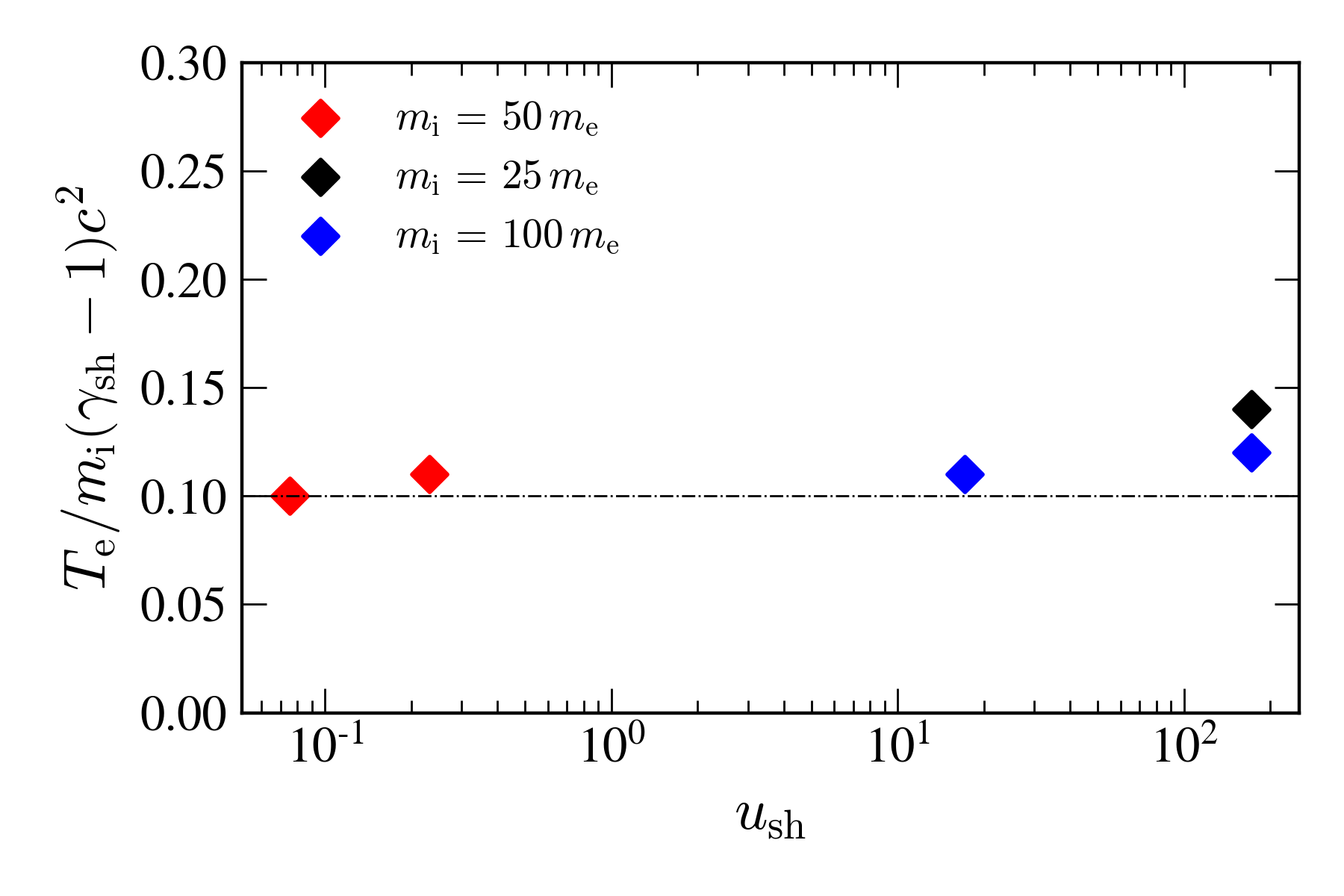}
    \caption{Electron temperature across the transrelativistic regime as obtained from PIC simulations for velocities listed in Tab.~\ref{tab:simulations}. The thermal energy imparted to electrons invariably accounts for about 10\% of the incoming kinetic energy in the shock front frame $\mi \left( \gamma_{\rm sh} - 1\right) c^2$.}
    \label{fig:TvsUsh}
\end{figure}

Our simulations are restricted to the limit of vanishing magnetization. In this regime, the shock is mediated by the Weibel instability~\cite{Weibel_1959, Medvedev_1999, Lemoine_2019_PRL, Vanthieghem_2024}. The Weibel instability is magnetically dominated -- i.e., $E^2 - B^2\,<\,0$ -- and grows at kinetic scales from the relative drift between the ambient plasma and reflected particles. As displayed in Table~\ref{tab:simulations}, the downstream electron temperature remains close to $k_{\rm B} T_e \,\simeq\, 0.1 \left( \gamma_{\rm sh} - 1\right) \mi c^2$ up to a very good approximation over the full set of transrelativistic shock velocities. While motivating a constant specific value for the downstream electron temperature is the topic of the following sections, we can already elaborate on the value of the temperature ratio discussed in the literature. Non-relativistic shocks show $T_{\rm e}/T_{\rm i}\,~\sim\,0.3$ in the high $M_{\rm A}$ regime~\cite{Kato_2008, Kato_2010, Vanthieghem_2024}, the fully relativistic regime tends to point towards larger temperature ratios $T_{\rm e}/T_{\rm i}\,~\sim\,0.5$~\cite{Spitkovsky_2008, Martins_2009, Haugbolle_2011, Sironi_2013, Vanthieghem_2022}. In terms of jump condition for a fixed downstream electron temperature $k_{\rm B} T_e \,\simeq\, \alpha \left( \gamma_{\rm sh} - 1\right) \mi c^2$, the temperature ratio between electrons and ions takes the form: 
\begin{align}
    &\frac{T_{\rm e}}{T_{\rm i}}\,=\,\frac{9 \alpha}{4 - 9 \alpha}\,\simeq\,0.3 \quad{\rm for}\quad|\beta_{\rm sh}| \ll 1, \qquad \frac{T_{\rm e}}{T_{\rm i}}\,=\,\frac{6 \alpha}{\sqrt{3} - 6 \alpha}\,\simeq\,0.5 \quad{\rm for}\quad |\beta_{\rm sh}| \lesssim 1\,,
\end{align}
where the figure of value corresponds to $\alpha\,=\,0.1$. These estimates come from matching the total pressure inferred from the Rankine-Hugoniot jump conditions with the partition of the pressure between electrons and ions parametrized by $\alpha$. The increase in the temperature ratio in the relativistic regime occurs between $|u_{\rm sh}|\,\simeq\,[0.2, 10]$. These results, illustrated in Fig.~\ref{fig:TvsUsh}, seem to point toward some general fluid argument bounding from above the fraction of energy pumped from the electron distribution before shock formation, similar to fluid models for the beam-plasma coupling in the shock upstream~\cite{Lemoine_2019_II}. The origin of the temperature equilibration has been discussed separately in the Newtonian and relativistic regimes (see~\cite{Vanthieghem_2024, Vanthieghem_2022} and references therein). Here, we provide a unified model for the electron temperature in the transrelativistic regime.

\begin{table}[htbp]
    \centering
    \begin{tabular}{ccccc} 
        \hline
        Run & $\beta_{\mathrm{sh}}$ & $m_i$ & $k_B T_e$  \\
            &                        & $[\me]$ & $[\left(\gamma_{\rm sh} - 1 \right) \mi c^2]$ \\[0.2cm]
        \hline
        N.1 & $0.075$ & $49$  & $0.10$  \\
        N.2 & $0.225$ & $49$  & $0.11$  \\
        R.1 & $0.9983$ & $100$ & $0.11$  \\
        R.2 & $0.999983$ & $25$  & $0.14$  \\
        R.3 & $0.999983$ & $100$  & $0.12$  \\
        \hline
    \end{tabular}
    \caption{Parameters and measured electron temperature for the set of transrelativistic PIC simulations for velocity ranging between $\beta_{\rm sh}\,=\,[0.075,\, 0.999983]$, where $\beta_{\rm sh}\,=\,v_{\rm sh}/c$ is the shock velocity measured in the far upstream frame normalized to the speed of light. The proper temperature is normalized to the kinetic energy of the shock and remains close to $k_{\rm B} T_e \,\simeq\, 0.1 \left( \gamma_{\rm sh} - 1\right) \mi c^2$ across the full range of shock velocities up to a good approximation. The evolution of the proper temperature in terms of shock velocity is shown in Figure~\ref{fig:TvsUsh}.}
    \label{tab:simulations}
\end{table}

\section{Electron transport in a decelerating magnetic turbulence}

Transport of the particles is best studied in the comoving frame of the magnetic structures. A complete kinetic description of the scattering center frame is discussed in~\cite{Pelletier_2019}.  This frame, denoted $\Rw$ for the Weibel turbulence, drifts close to the electron velocity and decelerates towards the shock in the shock front frame, $\Rs$. In $\mathcal{R}_{\rm w}$, the microturbulence is assumed to have a dynamical time scale much larger than the typical scattering time of the particles and is, therefore, quasi-magnetostatic. Such a description has the advantage of disentangling the transverse motional and longitudinal electric fields in the equation of motion expressed in $\Rw$:
\begin{align}\label{eq:motion}
    \frac{ {\rm d} \mathbf{p}_{\rm | w} }{ {\rm d} t_{\rm |w}}\,=\, \mathbf{p}_{\rm | w} \cdot \delta \hat{\mathbf{\Omega}}_t + q\,\mathbf{E} - \mathbf{\Gamma}_{a b} \frac{p^a_{\rm | w} p^b_{\rm | w}}{p^t_{\rm | w}}\,
\end{align}
where $\mathbf{p}_{\rm | w} \cdot \delta \hat{\mathbf{\Omega}}_t$ accounts for pitch-angle variations in the magnetostatic field, $\mathbf{E}\,=\,(E_x, 0, 0)$ is the purely longitudinal component of the electric field, and the connections $\Gamma^i_{a b}$ [$i=x, y, z$; $(a,b) = t, x, y, z$] account for inertial correction associated with the deceleration of $\Rw$. The non-vanishing stationary components of $\mathbf{\Gamma}_{ab}$ are $\Gamma^t_{tx} \,=\, \Gamma^x_{tt} \,=\, \partial_x \gamma_{\rm w}$ and $\Gamma^t_{xx} \,=\,\Gamma^x_{xt} \,=\, \frac{1}{\beta_{\rm w}}\partial_x \gamma_{w}$, where $u_{\rm w}\,=\,\beta_{\rm w} \gamma_{\rm w}$ is the 4-drift velocity of $\mathcal{R}_{\rm w}$ in the shock front frame. We decompose the infinitesimal scattering operator on the rotation matrices~\cite{Plotnikov_2011}.  In the following, we assume $\delta\hat{\Omega}_t$ to be a Gaussian stochastic process. Hence, our approach becomes semi-dynamical. 
The above Langevin equation can be written in terms of variables $(p^x,p^y,p^z)\,\to\,(p,\mu=\cos\theta,\phi)$ 
where $\mu$ is the pitch-angle cosine. The stochastic differential equation for the norm of the momentum then reduces to
\begin{equation}
\dot{p}\,=\, \mu \, \left[  q E^x -  \gamma_{\rm w}^3 p^t \left( \mu \frac{p}{p^t} + \beta_{\rm w} \right)  \partial_x \beta_{\rm w} \right]\,.
\end{equation}
This decomposition allows to isolate the stochastic contribution in pitch-angle variation for the various geometries
$\dot{\theta}_{\rm st}\,=\, \chi_t\, \text{(2D)}$ and $\dot{\mu}_{\rm st}\,=\, \frac{p^y}{p} \chi^y_t + \frac{p^z}{p} \chi^z_t\, \text{(3D)}$. This leads to a straightforward interpretation in terms of the usual scattering operators of the associated transport 
equation $D_{\theta\theta}^{(2D)}\,=\,\nu$ and $D_{\mu\mu}^{(3D)}\,=\,\nu\,\left( 1 - \mu^2\right)$~\cite{Risken}, in respective 2D and 3D dimensions. 

The scattering frequency, $\nu\,=\,\Delta \alpha^2/\Delta t$, depends on the turbulent magnetization in the shock transition $\epsilon_B\,=\,B^2/4\pi n_\infty\mi \left( \gamma_{\rm sh} - 1\right)c^2$ normalized to the incoming kinetic energy of the shock, the structure of the turbulence parallel ($r_\parallel$) and transverse ($r_\perp$) to the shock normal, and particle momentum $p_{\rm |w}$ in $\Rw$. One can directly distinguish particles trapped in magnetic structures from others. Trapped particles scatter off the longitudinal perturbations through the decoherence of their bounce motion. An estimate is obtained from the bounce frequency, $\omega_B$, of the particle exiting the structure with a typical deflection from the coherent gyration $\Delta\alpha\,=\,\omega_B r_\perp/v_{\rm th|w}$, where $\beta_{\rm th|w}$ is the thermal velocity, over the transit time in the filament $c\Delta t\,\sim\,r_\parallel/v_{\rm th|w}$. Conversely, particles of gyroradius $\wi r_{\rm g}/c\,=\,\epsilon_B^{-1/2} p_{\rm |w}/[\mi (\gamma_{\rm sh} - 1)^{1/2} c]$, much larger than the typical scale of the turbulence -- \emph{i.e.}, untrapped -- are well approximated by small-angle scattering through $\Delta \alpha^2\,\sim\,r_\perp^2/r_{\rm g}^2$ and $c\Delta t\,\sim\, r_\perp/v_{\rm |w}$. The transition between scattering regimes occurs when the momentum of the particle becomes lower than $p_{\rm 0|w}\,\simeq\,\mi\left(\gamma_{\rm sh} - 1\right)^{1/2}\,\epsilon_B^{1/2} r_\perp\wi$, such that the Larmor radius of the particle is comparable to the size of the turbulence. The scattering frequency  in both regimes is therefore:
\begin{align} 
&\nu^{\rm trapped} \,=\, 2 \pi \frac{r_\perp}{r_\parallel} \epsilon_B^{1/2} \left ( \frac{p_{\rm |w}}{m_{\rm i} \left( \gamma_{\rm sh} - 1 \right)^{1/2} c} \right)^{-1} \beta_{\rm | w}\,\omega_{\rm pi}  \qquad\text{for}\quad p_{\rm |w}<p_{\rm 0|w}\,,\label{eq:scatt1}\\
&\nu^{\rm untrapped} \,=\, \frac{\omega_{\rm p} r_\perp}{c} \epsilon_B \left ( \frac{p_{\rm |w}}{m_{\rm i} \left( \gamma_{\rm sh} - 1 \right)^{1/2} c} \right)^{-2} \beta_{\rm | w}\,\omega_{\rm pi} \qquad\text{for}\quad p_{\rm |w}\geq p_{\rm 0|w}\,.\label{eq:scatt2}
\end{align}
We have seen that a transition between the two regimes should occur when $r_{\rm g |w}\,\sim\,r_\perp$. Therefore, matching the solution at $p_{\rm 0|w}$ imposes the reasonable rescaling of the scattering frequency by a constant of the order $2 \pi r_\perp/r_\parallel \,\sim\, \mathcal{O}(1)$. Kinetic simulations suggest $\epsilon_B\,\sim\,10^{-2} - 10^{-3}$ over the shock transition and $\omega_{\rm pi} r_\perp/c\,\gtrsim\,10$.

The electric field originates from the difference in inertia between electrons and ions. Rapid pitch-angle scattering isotropizes the distribution accelerated by the ambipolar electric field. Trapping of the thermal electrons is ensured over the shock transition. It is, therefore, natural to analyze their dynamics in the fully diffusive regime -- i.e., $f_e(p,\mu)\,\simeq\, f_e^0(p) + \mu f_e^1(p)$. Following the same approach as in~\cite{Krimskii_1981, Vanthieghem_2022, Vanthieghem_2022b} for the transport equation associated with Eq.~\eqref{eq:motion}, we obtain a Fokker-Planck equation for the electrons distribution:
\begin{align}\label{eq:FP}
    \beta_{\rm w}\,\partial_x f - \frac{1}{3}\,p_{\rm |w}\,\frac{\partial_x u_{\rm w}}{\gamma_{\rm w}}  \partial_{p_{\rm | w}} f\,=\, \frac{1}{3 p_{\rm |w}^2} \partial_{p_{\rm | w}}  \left[ p_{\rm |w}^2 \frac{\left( q E_x - \sqrt{m^2 + p_{\rm |w}^2}\,\beta_{\rm w} \partial_x u_{\rm w} \right)^2}{\gamma_{\rm w} \nu} \partial_{p_{\rm | w}} f \right]
\end{align}
where we neglected spatial diffusion associated with particle injection and acceleration, see for instance~\cite{Krimskii_1981, Amano_2020, Grassi_2023}. In the above equation, the electric field amplitude is set by the incoming kinetic energy such that the electron momentum diffusion coefficient $D_{pp}\,\sim\,\tfrac{1}{3}q^2 E_x^2/\nu$ as long as $\tfrac{\mi}{\me}\gg\gamma_{\rm e|w}$. Analytical estimates of the energy partition between electrons and ions can then be obtained in the fluid regime~\cite{Vanthieghem_2024}.

\begin{figure*}[ht]
	\begin{center}
	\includegraphics[width=1.\textwidth]{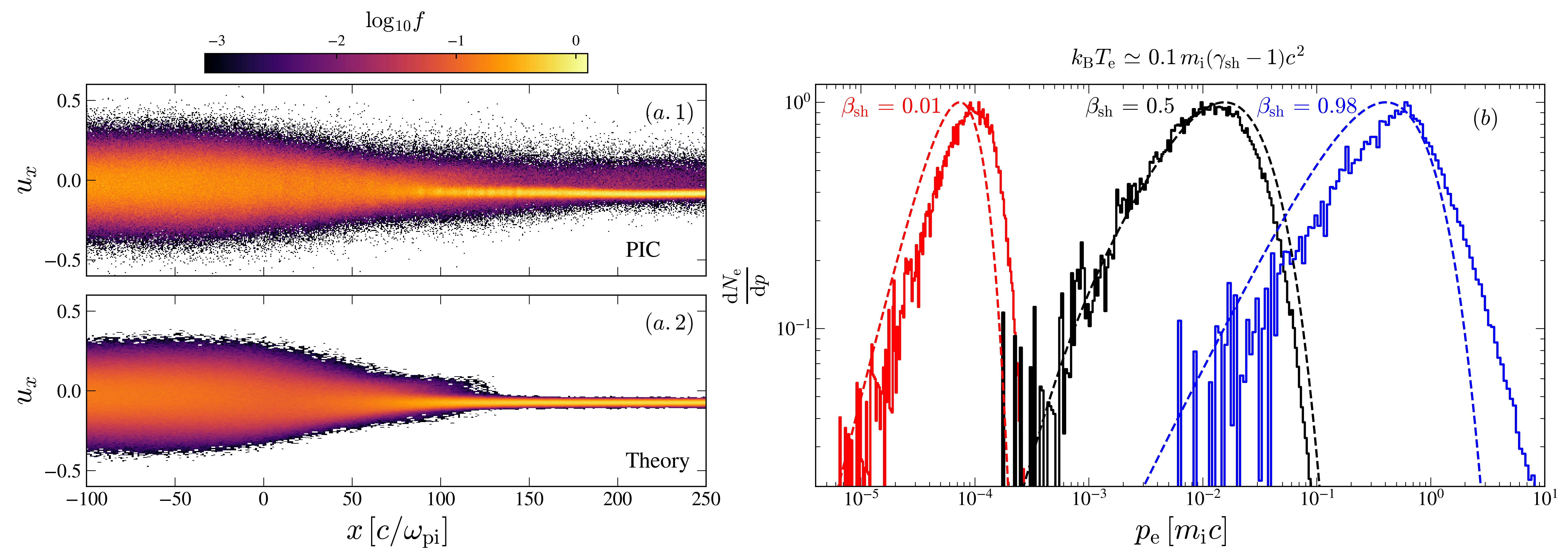}
	\caption{(a) Comparison of the electron $x-u_x$ phase space distribution between the PIC simulations (a.1-2) and integration of the Langevin equation~\eqref{eq:motion}. The shock is nonrelativistic with $\beta_{\rm sh}\,=\,0.075$ with parameters N.1 in Tab.~\ref{tab:simulations}. The figure is adapted from~\cite{Vanthieghem_2024}. (b) Solution of the Fokker-Planck equation~\eqref{eq:FP} for various shock velocities $\beta_{\rm sh}\,=\,0.01,\,0.5,\,0.98$ for $\epsilon_{B}\,=\,10^{-3}$, $\Delta \phi\,\sim\,\mi\left( \gamma_{\rm sh} - 1\right) c^2$, $\wi r_\perp/c\,\sim\,10$, $\wi L_{\rm sh}/c\,=\,150$, and pitch-angle scattering frequency are taken from Eqs.~\eqref{eq:scatt1}-\eqref{eq:scatt2}. Dashed lines show the comparison with the associated Maxwellian distribution of temperature $k_{\rm B} T_{\rm e}\,=\,0.1\,\mi\left(\gamma_{\rm sh} - 1\right)c^2$. }
	\label{fig:FP}
	\end{center}
\end{figure*}

A full solution of the transport equation, Eq.~\eqref{eq:motion}, can be found by integrating self-consistently the electrostatic field in the shock front frame using a Particle-In-Cell Poisson method. The longitudinal electric field is obtained from the charge density of the electrons and ions in the shock-front frame. The equation of motion of both species is solved in $\Rw$ using the invariance of $E_x$ by boosting from $\Rs$ to $\Rw$. To ensure the stability of the method, we impose a fixed $\Delta t_{|\rm sh}$, such that for each particle in $\Rw$, we have $\Delta t_{|\rm w}\,=\,\gamma_{\rm w}\,\left(\beta_{\rm w} \beta^x_{\rm | w} + 1 \right) \,\Delta t_{|\rm sh}$. The contribution of the electric field, non-inertial components, and pitch-angle scattering are then applied using a second-order Strang splitting~\cite{Strang_1968}. Good agreement is found between the run N.1 and the solution of the full transport equation~\eqref{eq:motion} as shown in panel (a.1-a.2) in Fig.~\ref{fig:FP}. 

Albeit not self-consistently coupling electrons and ions, direct integration of the Fokker-Planck equation, Eq.~\eqref{eq:FP}, gives a direct estimate of the electron temperature at the cost of assuming a typical amplitude for the cross-shock potential $\Delta \phi\,\sim\,\mi \left(\gamma_{\rm sh} - 1\right) c^2$. Panel (b) in Fig.~\ref{fig:FP} shows the electron spectra obtained from integration of Eq.~\eqref{eq:FP} for a large range of shock velocities $\beta_{\rm sh}\,=\,0.01,\,0.5,\,0.98$ using typical parameters. In each case, integration recovers efficient electron heating up to $k_{\rm B} T_{\rm e}\,=\,0.1\,\mi\left(\gamma_{\rm sh} - 1\right)c^2$.

\section{Conclusion}

Our results show that the electron temperature systematically accounts for about $\sim 10\%$ of the incoming kinetic energy of the associated weakly magnetized shocks over a large range of shock velocities. We studied energy partition and the physical mechanism at play through large-scale PIC simulations, a semi-analytical transport model from which we inferred a reduced analytical Fokker-Planck description.
In terms of temperature ratio, this fraction translates into an increase from the non-relativistic shocks of typical $T_{\rm e}/T_{\rm i}\,\sim\,0.3$ to relativistic ones with $T_{\rm e}/T_{\rm i}\,\sim\,0.5$. 
From the non-relativistic to the ultra-relativistic regime, our model gives a dominant and unified mechanism for the temperature equilibration between electrons and ions. Electrons effectively scatter in pitch-angle through fast decoherence of their bounce motion in the magnetized structures. The difference in scattering frequency between electrons and ions generates an ambipolar electric field that accelerates the electrons, which are then continuously isotropized by fast pitch-angle scattering. Ambipolar electron heating is well captured in the diffusive regime, from which we derive a diffusion coefficient $D_{pp}\,\sim\,\tfrac{1}{3} e^2 E_x^2/\nu$. Finally, we have shown that integration of the associated Fokker-Planck equation naturally recovers the electron temperature observed in PIC simulations over a large range of shock velocities. 
In the context of binary-neutron-star mergers, such as GW 170817A~\cite{Abbott_2017a,Abbott_2017b}, our findings indicate a consistent fraction of kinetic energy imparted to thermal electrons throughout the entire afterglow phase. Together with the proper modeling of the long-term non-thermal spectral evolution~\cite{Takahashi_2022, Groselj_2024, Margalit_2022, Margalit_2024}, such effects still need to be further explored.

\bibliographystyle{JHEP}
\bibliography{bib_letter}

\end{document}